\documentclass[
reprint,
superscriptaddress,
 amsmath,amssymb,
 aps,
]{revtex4-2}

\usepackage{graphicx}
\usepackage{dcolumn}
\usepackage{bm}

\begin{document}

\author{Raphaël Rousset-Zenou}
\affiliation{University Grenoble Alpes, CNRS, Grenoble INP, Institut Néel, 38000 Grenoble, France}
\author{Nicolas Aparicio}
\affiliation{University Grenoble Alpes, CNRS, Grenoble INP, Institut Néel, 38000 Grenoble, France}\author{Simon Messelot}
\affiliation{University Grenoble Alpes, CNRS, Grenoble INP, Institut Néel, 38000 Grenoble, France}
\author{Rasmus D. Schlosser}
\affiliation{Center for Quantum Devices, Niels Bohr Institute, University of Copenhagen, 2100 Copenhagen, Denmark}
\author{Martin Bjergfelt}
\affiliation{Center for Quantum Devices, Niels Bohr Institute, University of Copenhagen, 2100 Copenhagen, Denmark}
\author{Julien Renard}
\affiliation{University Grenoble Alpes, CNRS, Grenoble INP, Institut Néel, 38000 Grenoble, France}
\author{Moïra Hocevar}
\affiliation{University Grenoble Alpes, CNRS, Grenoble INP, Institut Néel, 38000 Grenoble, France}
\author{Jesper Nygård}\email{nygard@nbi.ku.dk}
\affiliation{University Grenoble Alpes, CNRS, Grenoble INP, Institut Néel, 38000 Grenoble, France}
\affiliation{Center for Quantum Devices, Niels Bohr Institute, University of Copenhagen, 2100 Copenhagen, Denmark}

\title{Gate-tunable Josephson parametric amplifiers based on semiconductor nanowires}

\begin{abstract}
Superconductor-semiconductor hybrid materials have been extensively used for experiments on electrically tunable quantum devices. Notably, Josephson junctions utilizing nanowire weak links have enabled a number of new gate-tunable qubits, including gatemons,  Andreev level qubits and spin qubits. 
Conversely, superconducting parametric amplifiers based on Josephson junctions have not yet been implemented using nanowires, even though such nearly quantum limited amplifiers are key elements in experiments on quantum circuits. Here we present Josephson parametric amplifiers based on arrays of parallel InAs nanowires that feature a large critical current as required for linear amplification. The resonance frequency of the devices is gate-tunable by almost 1 GHz, with a gain exceeding 20 dB in multiple frequencies and markedly improved noise performance compared to a low-temperature commercial amplifier.  
This platform enables on-chip integration of gate-tunable qubits with nearly quantum limited amplifiers using the same hybrid materials and on any substrate.
\end{abstract}

\maketitle

\section{Introduction}
Superconducting Josephson parametric amplifiers \cite{zimmerParametricAmplificationMicrowaves1967} allow the addition of the minimum amount of noise permitted by quantum mechanics in the amplification process \cite{cavesQuantumLimitsNoise1982, clerkIntroductionQuantumNoise2010}. As such, they are now essential in quantum bit readout and microwave quantum optics experiments when the signal of interest is below the noise floor of standard cryogenic amplifiers. They allow for instance fast and accurate single-shot measurement of superconducting quantum bits and nano-mechanical resonators \cite{walterRealizingRapidHighFidelity2017,teufelSidebandCoolingMicromechanical2011} and vacuum noise squeezing \cite{castellanos-beltranAmplificationSqueezingQuantum2008}.
A typical superconducting parametric amplifier, when operated as a resonant amplifier, consists of a microwave cavity in which a tunnel Josephson junction is inserted. The presence of the Josephson junction will introduce non-linearity arising from its current-phase relation. As this amplification process relies on a resonant phenomenon, the bandwidth is intrinsically limited to a few MHz. New designs of flux-tunable \cite{castellanos-beltranWidelyTunableParametric2007, yamamotoFluxdrivenJosephsonParametric2008a} and wide-band traveling-wave parametric amplifiers \cite{macklinQuantumlimitedJosephsonTravelingwave2015, planatPhotonicCrystalJosephsonTravelingWave2020} have been created to circumvent this issue. 
Another approach that recently sparked a significant interest, is to introduce a semiconducting weak link as the Josephson junction. 
The presence of the semiconductor allows for its transport properties, such as non-linearity or critical current, to be tuned using the electric field generated by a gate voltage. Recent implementations of this idea have led to the development of gate tunable Josephson parametric amplifiers (JPAs) using graphene \cite{butseraenGatetunableGrapheneJosephson2022, sarkarQuantumnoiselimitedMicrowaveAmplification2022} and planar InAs \cite{haoKerrNonlinearityParametric2024, phanGatetunableSuperconductorsemiconductorParametric2023}.

Among hybrid superconductor-semiconductor (Sc-Sm) devices, InAs nanowires with epitaxial aluminum have been widely used in gate-tunable quantum bits \cite{pita-vidalNovelQubitsHybrid2025,larsenSemiconductorNanowireBasedSuperconducting2015, casparisGatemonBenchmarkingTwoQubit2016} as they provide a defect-free Sc-Sm interface \cite{krogstrupEpitaxySemiconductorSuperconductor2015} and the nanowire geometry allows for an easy transfer onto a substrate suitable for microwave circuits with low dielectric loss. Recently, kinetic inductance parametric amplifiers \cite{splitthoffGatetunableKineticInductance2024} and Andreev qubits have also been demonstrated with the same materials \cite{pita-vidalStrongTunableCoupling2024, cheungPhotonmediatedLongrangeCoupling2024, haysCoherentManipulationAndreev2021}. Additionally hybrid nanowires have been of great interest in search for new bound states and topological features in quantum devices \cite{pradaAndreevMajoranaBound2020}. 

In a Josephson junction, the amount of non-linearity arising from the current-phase relation usually scales inversely with the critical current of the junction. While one-dimensional semiconductor nanowire weak links can provide a large non-linearity (small critical current), suitable for realizing quantum bits, this large non-linearity is a challenge in building a parametric amplifier. Indeed, the power-handling capabilities of such devices are expected to decrease with increasing non-linearity \cite{eichlerControllingDynamicRange2014}. This explains why nanowires with low critical currents have not been used in Josephson parametric amplifiers yet.

We demonstrate here JPAs based on Josephson junctions constituted of parallel InAs semiconducting nanowires that feature a large critical current and a reduced non-linearity, not hitherto realized with single nanowires. By integrating the Josephson junctions in a microwave resonator, we demonstrate gate-tunability of the resonance frequency by almost 1 GHz and, in optimal conditions, a parametric gain exceeding 20 dB. The amplifier has a limited compression point, typical in such single junction resonant amplifiers. We show that using this amplifier improves the signal-to-noise ratio of the measurement chain compared to a standard cryogenic amplifier. 

\begin{figure}
	\centering
	\includegraphics[width=0.9\columnwidth]{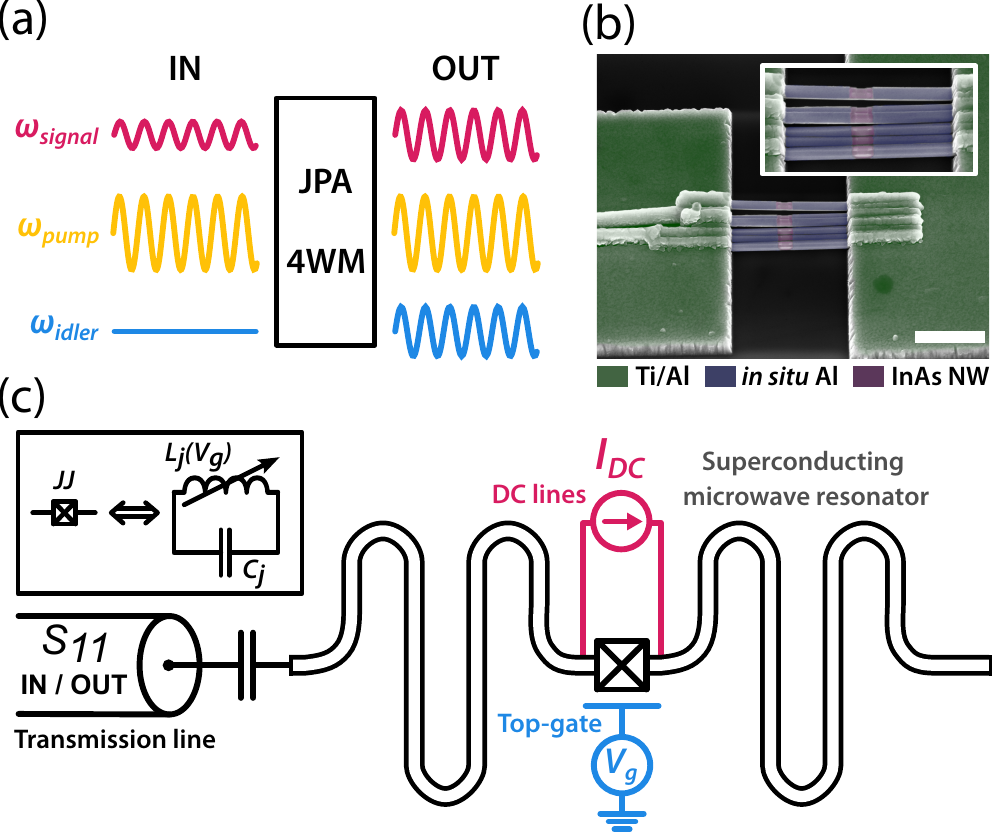}
	\caption{\textbf{Elements of the nanowire-based parametric amplifier} (a) Schematic of the working principle of a four-wave mixing Josephson parametric amplifier (b) False colored scanning electron micrograph of the parallel nanowire Josephson junction. 
		Ti/Al electrodes are used to contact the nanowires on which Al has been deposited in situ after the nanowire growth. Part of the aluminum shell is chemically etched, yielding a parallel nanowire Josephson junction.
		Scale bar 2 $\mu$m (c) Schematic of the JPA device with a junction embedded in an Al microwave resonator. $I_{DC}$ denotes the dc current sent to junction through the dc lines. $V_g$ denotes the gate voltage applied to the Josephson junction. Inset top-left: linear equivalent circuit of the semiconducting Josephson junction. The inductance of the junction is gate tunable.}
	\label{fig:device_schematic}
\end{figure}
\section{Device design}
Arrays of nanowires are grown parallel to each other in order to increase the number of conducting channels (and thus the critical current) while maintaining gate tunability (see Supplemental Material (SM) 
section I for pictures of as-grown arrays \cite{Note1}\footnotetext{See Supplemental Material [url] for detailed device fabrication procedure, device images, experimental measurement setup, measurement of current-phase relationship, noise analysis.
}). 
To achieve a pristine interface, an aluminum half-shell is subsequently deposited on the nanowires in situ and at low temperature. Part of the aluminum shell is etched away chemically, resulting in a Josephson junction (JJ) with a weak link composed of parallel semiconducting nanowires shown in Fig. \ref{fig:device_schematic}b.  
Two different devices were fabricated from the same batch of nanowires using the same nano-fabrication process (SM section I for materials, images and fabrication process \cite{Note1}). The first device (JPA04) is composed of an array of 4 parallel nanowires and the second device (JPA09) features three arrays for a total of 12 nanowires. 
The parallel nanowire Josephson junction is embedded at the center of a $\lambda/2$ microstrip resonator (Fig. \ref{fig:device_schematic}c) with a simulated bare resonance frequency $\omega_0/2\pi$ = 6.5 GHz (see Appendix \ref{App:design} and SM for details on resonator design \cite{Note1}). The resonator is capacitively coupled to a 50 $\Omega$ transmission line which is used to probe the reflection scattering parameter $S_{11}$ of the device. Low frequency lines are galvanically connected to the Josephson junction, near the center of the resonator, in order to perform DC transport measurements. A top-gate is used to tune the carrier density in the semiconducting nanowires. The schematic in Fig. \ref{fig:device_schematic}a illustrates how the design of the resonator allows to implement four-wave mixing (4WM) parametric amplification with the relation $2\omega_{\text{pump}} = \omega_{\text{signal}} + \omega_{\text{idler}}$. This corresponds to the annihilation of two pump photons at $\omega_{\text{pump}}$ to create a photon at $\omega_{\text{signal}}$ and a photon at $\omega_{\text{idler}}$. 

\section{Tunability}
\begin{figure}
	\centering
  \includegraphics[width=0.8\columnwidth]{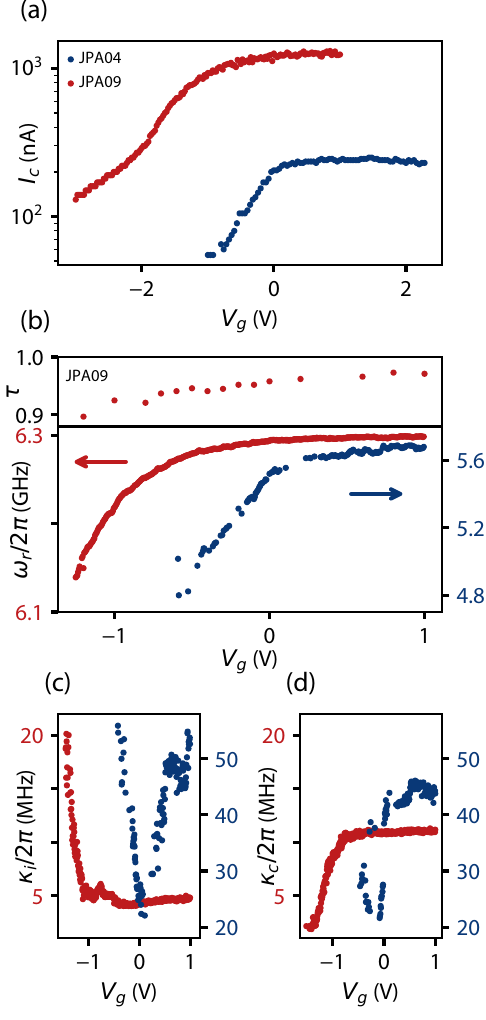}
	\caption{\textbf{Gate-tunability of the microwave resonator} (a) Measured critical current with respect to the applied gate voltage on the nanowires in device JPA04 (blue) and JPA09 (red). (b) Evolution of the resonance frequency of the device with respect to the gate voltage.  Effective transparency $\tau$ of the Josephson junction as a function of the gate voltage for device JPA09. (c) Internal loss rate $\kappa_i$ and (d) coupling rate $\kappa_c$ as a function of gate voltage $V_g$ for both devices.}
	\label{fig:spectro}
\end{figure}

The dc properties of the Josephson junctions are presented in Fig. \ref{fig:spectro}a. 
The critical current $I_c$ is extracted from a differential resistance measurement as a function of the dc bias current and the gate voltage. 
The top-gated nanowire junction acts as a Josephson field-effect transistor. 
$I_c$ saturates towards positive $V_g$ values while a negative voltage results in a dramatic reduction of $I_c$ towards pinch-off. 
The critical current in the parallel nanowire junctions is higher than for a single InAs nanowire/aluminum device (typically $<$ 50 nA) \cite{zellekensHardGapSpectroscopySelfDefined2020, nishioSupercurrentInAsNanowires2011, yangProximitizedJosephsonJunctions2021} showcasing the usefulness of the parallel arrays in achieving the necessary high critical current. 
Furthermore, by placing several arrays in close proximity and contacting them in parallel, the critical current can be further increased to more than 1200 nA in device JPA09.

We now turn to the high frequency regime. Sending a microwave probe tone allows to extract the resonance frequency $\omega_r$ of the device (see Methods for details on the measurement). The resonance frequency can be modulated by applying a gate voltage on the nanowires (Fig. \ref{fig:spectro}b) and devices JPA04 and JPA09 are tunable by more than 800 MHz and 150 MHz, respectively.
In device JPA09, when the gate voltage is set to obtain a large critical current, the resonance frequency of the device is close to the bare resonance frequency of the resonator $\omega_0$. However, increasing $V_g$ until saturation of device JPA04 results in a resonance frequency 800 MHz below $\omega_0$, a direct consequence of the smaller critical current in this device.
In a simple picture the Josephson junction acts as a gate tunable inductor (inset in Fig. \ref{fig:device_schematic}c) and thus affects the resonance frequency of the resonator $\omega_r$.
We consider the Josephson inductance $L_J$,
\begin{equation}\label{eq:Lj}
	L_J = \dfrac{\Phi_0}{2\pi}\left(\dfrac{\partial I}{\partial \phi}\right)^{-1} \approx \dfrac{\Phi_0}{2\pi}\left(\dfrac{1}{I_c}\right)
\end{equation} 
with $\Phi_0 = \dfrac{h}{2e}$, the superconducting flux quantum, and $\phi$ is the superconducting phase difference. The second equality is only correct in the limit of a sinusoidal current phase relation (CPR), $I=I_c\sin(\phi)$,  near $\phi=0$. For a small microwave probe power, it is reasonable to consider $\phi \approx 0$.
In Josephson junctions utilizing a semiconducting weak link with a large transparency, this CPR has been shown to not be accurate \cite{golubovCurrentphaseRelationJosephson2004}. Nevertheless, it allows to grasp the behavior of the resonator frequency when tuning the critical current via the gate voltage. When the critical current $I_c$ is large, $L_J$ is small (Eq. \ref{eq:Lj}). 
The resonance frequency $\omega_r$ of the resonator will therefore be dominated by the geometric inductance $L_0$ of the microstrip resonator and its effective capacitance $C$: 
\begin{equation}\label{eq:tunability}
	\omega_r(V_g) = \left[ \sqrt{(L_0 + L_J(V_g))C} \right]^{-1}
\end{equation}

When lowering $I_c(V_g)$, $L_J$ will increase, decreasing $\omega_r$. This explains the difference in resonance frequency of both devices, even at the same $V_g$.
The effective transparency $\tau$ \cite{blonderTransitionMetallicTunneling1982} of the Josephson junction can be extracted by an indirect measure of the CPR. Biasing the junction with a dc current, the resonance frequency of the device changes with the evolution of the phase across the junction. In other terms, one can access the Josephson inductance, i.e. the local derivative of the CPR and thus reconstruct the CPR of the junction (SM section IV \cite{Note1}). The evolution of $\tau$ with gate voltage is presented in  Fig. \ref{fig:spectro}b. It is close to unity, which suggests low disorder in the nanowire and at the interface between the InAs nanowires and the aluminum. Such high transparency was already reported in similar systems \cite{goffmanConductionChannelsInAsAl2017, zellekensHardGapSpectroscopySelfDefined2020}. This will result in a skewed current-phase relation with a reduced non linearity compared to tunnel junctions made of Al/AlOx \cite{spantonCurrentPhaseRelations2017}. The consequences of a non-sinusoidal CPR for the performance of the amplifier are discussed below. 

\section{Microwave losses and non-linearities}

By fitting the microwave reflection following Ref. \cite{probstEfficientRobustAnalysis2015}, we extract the internal loss rate $\kappa_i$ (Fig. \ref{fig:spectro}c) and the coupling rate $\kappa_c$ (Fig. \ref{fig:spectro}d) as a function of $V_g$. Extracting these quantities is critical to understand the behavior of the device and to ensure that the noise performance is not limited by internal losses, i.e.\ that we are in a regime  $\kappa_c > \kappa_i$. A sharp increase in internal losses is observed at negative gate voltages while the coupling rate decreases around the same gate voltage values, for both devices. While the internal losses can be mainly attributed to dissipation in the junction \cite{phanGatetunableSuperconductorsemiconductorParametric2023}, the coupling rate is dominated by the geometrical coupling capacitance to the transmission line. In Fig. \ref{fig:spectro}c, internal losses are large even at positive $V_g$ for device JPA04 and decrease towards zero gate voltage. This could indicate that, in this particular device, dissipation is induced by the gate itself. We do not currently understand the microscopic mechanism for the dissipation in the devices. On-chip filtering on the gate or a thicker dielectric could reduce this parasitic phenomenon. Also, the precise gate dependence of the external coupling is not the one expected from simple consideration on the evolution of the resonance frequency and might be due to the existence of spurious microwave modes.   
The net effect of internal dissipation is to restrain the useful amplification range to a smaller frequency band where losses are not predominant. 

\begin{figure}
    \centering
  \includegraphics[width=0.9\columnwidth]{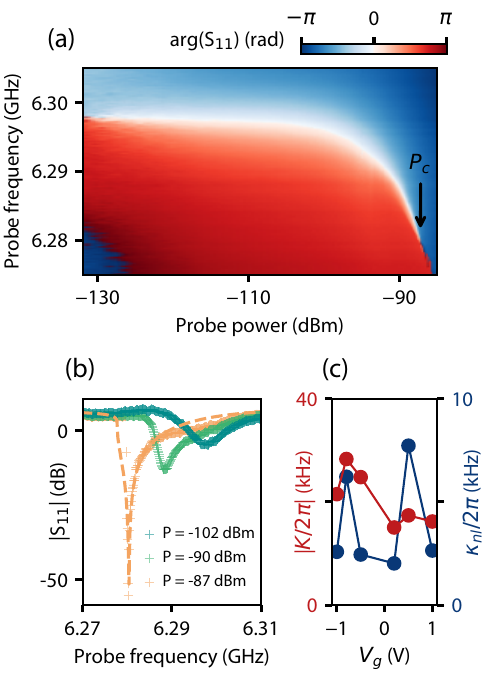}
    \caption{\textbf{Kerr non-linearity of the resonator with nanowire Josephson junctions} (a) Effect of the input power on the resonance frequency. Black arrow shows the bifurcation point at $P$ = $P_c$ (b) Magnitude of the microwave reflection with respect to frequency for various input powers $P$ at $V_g$ = 1V. Dashed line is a fit using the model described in the main text. 
    	(c) Kerr coefficient $K$ and non-linear loss rate $\kappa_{nl}$ dependence on the gate voltage, extracted by fitting $S_{11}$ at the power $P = P_c$ (All data from JPA09).}
    \label{fig:kerr}
\end{figure}

The presence of the Josephson junction adds non-linearity to the microwave resonator which can be studied by varying the input probe power (SM section II \cite{Note1}). In Fig. \ref{fig:kerr}a, the evolution of the resonance is apparent and shifts towards lower frequency when increasing power. This is an evidence of a Kerr-like non-linearity, a critical ingredient to perform parametric amplification \cite{manucharyanMicrowaveBifurcationJosephson2007}. 
The behavior of the device can be modeled by expanding the simplified non-linear Josephson CPR $\sin(\phi) \approx \phi - \phi^3/6$,  leading to a non-linear Kerr Hamiltonian \cite{rhoadsImpactNonlinearityDegenerate2010} 
\begin{equation}\label{eq:hamiltonian}
	H/\hbar = \left(\omega_r + \dfrac{K}{2}\hat{A}^{\dagger}\hat{A} \right) \hat{A}^{\dagger}\hat{A}
\end{equation}
with $A^{\dagger}$($A$), the creation (annihilation) operator for the photons inside the resonator and $K$, the Kerr coefficient, quantifying the amount of non-linearity.\\
The (negative) Kerr term is a direct consequence of the non-linear Josephson potential and will cause a monotonic decrease of the resonance frequency, dependent on the number of photons in the resonator.
Above a critical input power $P_c$ (Fig. \ref{fig:kerr}b, $P_c$ = -87 dBm), a bifurcation occurs and the device enters a bistable regime where linear amplification is no longer possible \cite{vijayInvitedReviewArticle2009}. The largest gains are expected by setting the pump just below the bifurcation threshold. 
Taking line-cuts from Fig. \ref{fig:kerr}a at various probe powers shows the increase of the resonance dip, while it shifts to lower frequencies due to the Kerr non-linearity (Fig. \ref{fig:kerr}b). This is due to the presence of non-linear losses and has also been observed in graphene Josephson junctions \cite{schmidtBallisticGrapheneSuperconducting2018,butseraenGatetunableGrapheneJosephson2022} but is usually not observed in tunnel junctions \cite{boakninDispersiveMicrowaveBifurcation2007}. The losses are attributed to the dense spectrum of Andreev bound states within the induced superconducting gap of the junction, leading to dissipation in the junction \cite{schmidtProbingCurrentphaseRelation2020, hallerPhasedependentMicrowaveResponse2022}. Such losses are an indication of the coupling to other modes. The vacuum noise from these extra modes gets added to the noise of the amplifier, increasing its total noise and potentially preventing quantum-limited behavior.

We fit the reflection coefficient (dashed lines in Fig. \ref{fig:kerr}b) using the model developed by Yurke and Buks \cite{yurkePerformanceCavityparametricAmplifiers2006} as it takes into account non-linear losses. It allows to extract the Kerr coefficient ($K$), whose evolution as function of gate voltage is presented in Fig. \ref{fig:kerr}c for device JPA09 along with the non-linear losses $\kappa_{nl}$. A weakly non-linear regime is required for parametric amplification such that $K < \kappa_c$ \cite{bourassaJosephsonJunctionembeddedTransmissionline2012} and typically $|K|/\omega_0$ $<$ $10^{-2}$ to $10^{-6}$.
Kerr non-linearity is extracted by performing the fit at the critical power $P_c$ (SM section III for details \cite{Note1}) and decreases with increasing $V_g$ (Fig. \ref{fig:kerr}c) which is typical of JPAs based on semiconducting weak-links \cite{butseraenGatetunableGrapheneJosephson2022, phanGatetunableSuperconductorsemiconductorParametric2023,Samanta2026} as $K$ scales with $1/I_c$.
The value of $K/\omega_0 \sim 10^{-5}$ is significantly smaller than JPAs based on single tunnel JJs and similar to JPAs implemented with a semiconducting weak-link \cite{haoKerrNonlinearityParametric2024}. 
It has been shown that the Kerr non-linearity derived from the Josephson potential is linearly reduced with increasing transparency of the Josephson junction \cite{fatemiNonlinearityTransparentSNS2025}. As a small Kerr coefficient is expected to increase the power handling capabilities of the amplifier \cite{eichlerControllingDynamicRange2014} it can be concluded that a highly transparent junction is desirable in a JPA.

\section{Parametric amplification}
\begin{figure}
    \centering
   \includegraphics[width=0.9\columnwidth]{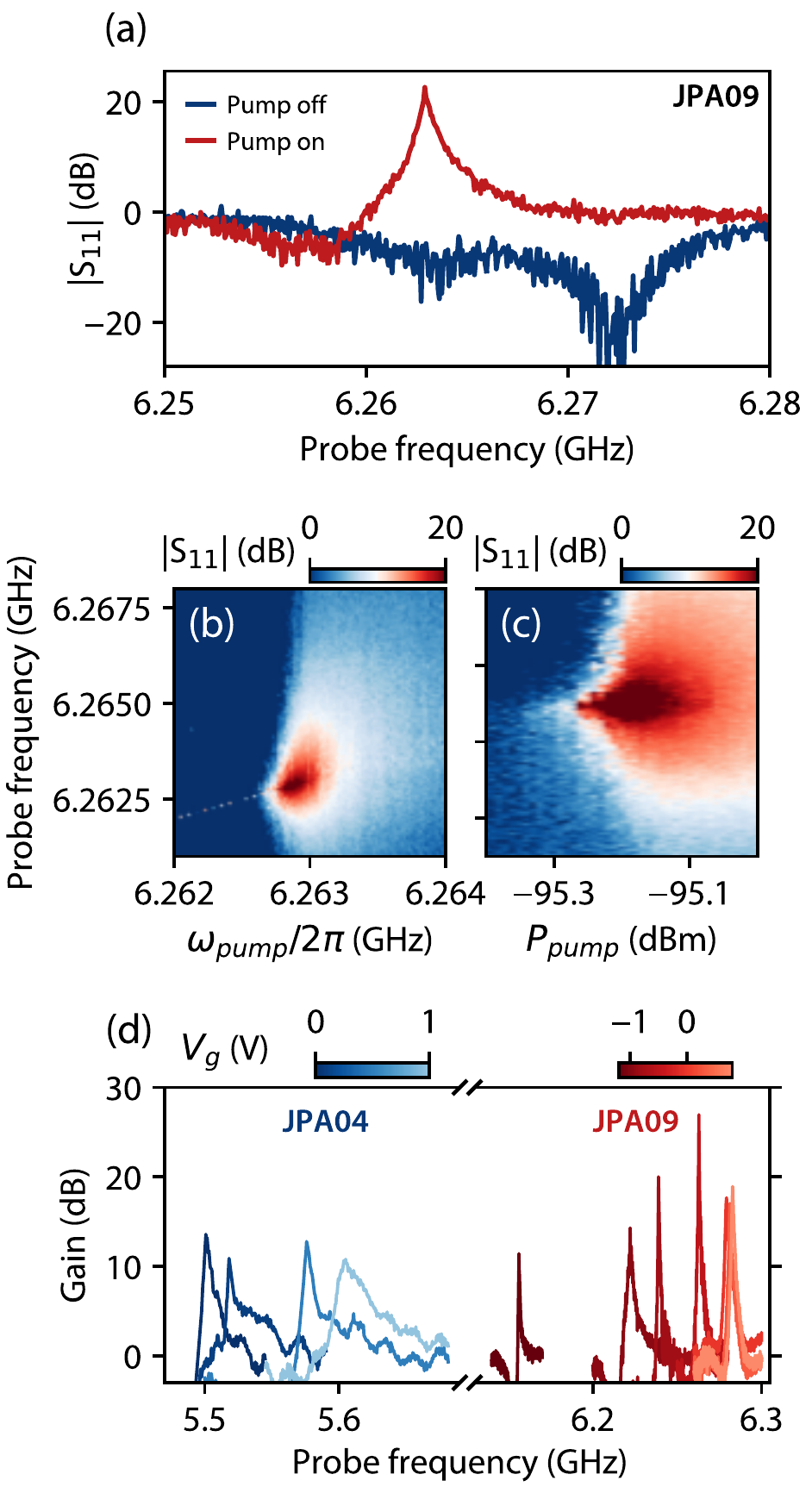}
	\caption{\textbf{Parametric amplification with the JPA} (a) Magnitude of the microwave reflection $S_{11}$ from device JPA04 when the microwave pump is turned off (blue line) and on (red). (b, c) Effect of the pump frequency ($\omega_{\text{pump}}/2\pi$) and power ($P_{\text{pump}}$) on the gain ($V_g$ = -0.5 V). (d) Effect of the gate voltage on the frequency tuning of the JPA. Traces taken at various gate voltages for device JPA04 (blue) and JPA09 (red).}
    \label{fig:gain_traces}
\end{figure}

We now present parametric amplification capabilities of the JPA. A strong microwave pump tone with a frequency $\omega_p$ is sent through the same port as the weaker probe tone of frequency $\omega_s$ while measuring the microwave reflection $S_{11}$. The microwave pump tone is periodically modulating the non-linear Josephson inductance leading to parametric gain. Fig. \ref{fig:gain_traces}a shows the effect of the pump on the device. When the pump is off, a resonance dip in magnitude is present. When the pump is turned on and correctly tuned in frequency and power, the reflected signal amplitude increases, evidence of parametric amplification. A Kerr-induced frequency shift of the resonance is observed due to the high power of the pump tone. \\
A careful optimization of the pump frequency (Fig. \ref{fig:gain_traces}b) and power (Fig. \ref{fig:gain_traces}c) is necessary and the maximal gain is obtained in a narrow parameter range.
In our devices, the instantaneous-bandwidth is in the range of a few MHz and set by the external quality factor of the resonator. We are able to tune the resonance frequency of the device using the gate voltage, and by modifying accordingly the pump power and frequency, we are able to perform parametric amplification in a range of more than 100 MHz as shown by the gain profiles in Fig. \ref{fig:gain_traces}d.  
A key metric of an amplifier is its dynamic range. Quantum-limited amplifiers with high dynamic range are of interest as they allow to increase the number of simultaneous qubit readouts, which is essential for scalability \cite{whiteReadoutQuantumProcessor2023, kaufmanJosephsonParametricAmplifier2023}.
The 1 dB compression point $P_{\text{1dB}}$ is defined as the input probe power at which the gain is 1 dB lower than the low power gain (Fig. \ref{fig:noise}a). The device is operated such as to produce 20 dB of gain at the lowest input power. The gate voltage is set at $V_g$ = -0.5 V and we apply a signal detuned from the pump $\delta \omega/2\pi$ = 200 kHz where $\delta \omega = \omega_{\text{signal}} - \omega_{\text{pump}}$. Increasing the signal input power results in a reduction of the gain, giving $P_{\text{1dB}}$ = -137 dBm. This is a relatively low value, which is expected in such single junction 4WM JPA \cite{castellanos-beltranBandwidthDynamicRange2009}. There are several ways to increase the compression point. Firstly, it has been shown that the ratio $\kappa_c/K$ scales positively with the dynamic range \cite{eichlerControllingDynamicRange2014}, thus an improvement of the coupling between the resonator and the transmission line, which can be implemented at the nanofabrication stage, or a reduction of the Kerr coefficient would lead to a better compression point (see SM section VI for details \cite{Note1}). Secondly, when operating the amplifier at large gain, the amplified signal can become a non-negligible fraction of the pump power, affecting the amplifier behavior and reducing its compression point. This pump back-action phenomenon is typical of Kerr-based amplifiers. Typical strategies to improve the stiffness of the pump involve the use of double-pumping \cite{kamalSignaltopumpBackAction2009} or operation at a higher pump power \cite{malnouOptimalOperationJosephson2018}.
Another strategy would be to operate the device as a three-wave mixing amplifier \cite{frattiniOptimizingNonlinearityDissipation2018} by pumping via the gate electrode similar to flux pumping \cite{yamamotoFluxdrivenJosephsonParametric2008a}.

\section{Noise analysis}

\begin{figure}
	\centering
 	\includegraphics[width=0.9\columnwidth]{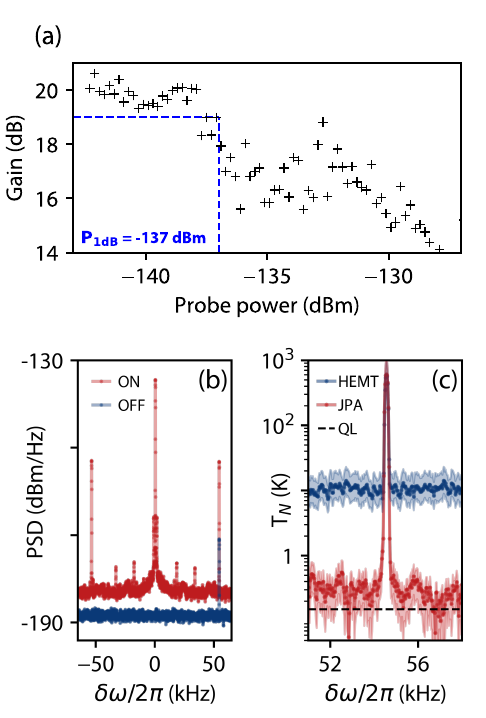}
	\caption{\textbf{Performance of the parametric amplifier} (a) Measurement of the 1dB compression point $P_{\text{1dB}}$. Blue dashed vertical line gives $P_{\text{1dB}}$ = -137 dBm. (b) Measured power spectral density with the pump turned off (blue line) and the pump turned on (red line) (c) System added noise temperature when the JPA is off (blue trace, pump is off) and when the JPA is on (red trace, pump is on). A relative reduction in noise is observed, while an uncertainty on absolute levels should be noted (see text). Error bars correspond to $\pm$ 2 dB uncertainty on the attenuation chain. The black dashed line (QL = 0.15 K at this frequency) corresponds to the quantum limit. (All data from JPA09)}
	\label{fig:noise}
\end{figure}

Low-noise Josephson parametric amplifiers, ideally reaching near quantum-limited performance, need to feature a sufficiently large gain (typically $\geq$ 20 dB) in order to overcome the noise of the following amplifier in the measurement chain. To evaluate the noise performance of the nanowire JPA, we assess the amplifier noise using a spectrum analyzer. The following measurements are done with device JPA09 that can feature a gain exceeding 20 dB (Fig. \ref{fig:gain_traces}). 
We first compare, in Fig. \ref{fig:noise}b, the effect of the pump on the measured power spectral density (PSD). The pump power and frequency ($P_{\text{pump}}$ = -95.5 dBm and $\omega_{\text{pump}}/2\pi$ = 6.26135 GHz) are tuned to obtain a gain of 22 dB. 
In the absence of the pump, we see the signal tone 18 dB above the noise floor. When the pump is turned on, we additionally observe the pump and idler tones. 

We also notice multiples smaller peaks whose origin is unclear at this stage. They may come from higher order non-linearities, neglected in Eq. \ref{eq:hamiltonian}, that arise from the different harmonics present in the CPR for JJs with a semiconducting weak link \cite{spantonCurrentPhaseRelations2017}, or more trivially from noise sources that appear when operating the resonator in the parametric amplifier regime, i.e. close to bifurcation. Additional data as a function of pump power is shown in SM sect. VI \cite{Note1}.

The noise floor is higher in the presence of the pump, due to the amplification of the noise by the JPA, but most importantly the signal to noise ratio, i.e.\ the difference between the signal and the noise floor has improved by 12 dB. This is the key feature of such amplifier, which proves that the noise temperature of the measurement chain has decreased thanks to the JPA operation.  

To be more quantitative, we translate in Fig. \ref{fig:noise}c the measured PSD into an effective added noise temperature $T_N$ (see SM  section V for details \cite{Note1}). When the pump is off, i.e. when the device shows no gain and the first amplifier of the chain is the HEMT thermalized to the 4~K stage of the fridge, we extract an added noise in the range 5 to 10 K, i.e. what is expected for such an amplifier. When the pump is turned on, the added noise temperature is significantly reduced. 
The theoretical quantum-limit is shown for comparison (black dashed line) \cite{cavesQuantumLimitsNoise1982}. 
However, we emphasize that this analysis is sensitive to the exact calibration of the chain and that a precise quantitative measurement of the noise performance would require a calibrated noise source at the input of the amplifier. Based on our experiments, it can only be concluded that the signal-to-noise ratio is significantly improved compared to the cryogenic HEMT amplifier.

\section{Conclusions and outlook}
In conclusion, we have demonstrated a gate-tunable Josephson parametric amplifier based on InAs/Al nanowires. To overcome the typically low critical current of such nanowires, unsuitable for realizing a parametric amplifier, we fabricated a device with several nanowires in parallel. The amplification band can be tuned using a gate voltage by more than 100 MHz and a gain over 20 dB can be achieved. 
We have shown that the Kerr nonlinearity of such nanowire junctions is reduced compared to tunnel junctions due to their large transparency. As this is a key parameter to optimize the performance of an amplifier, this could be leveraged in future designs. We envision that using three wave mixing \cite{buccheri2026kerrnonlinearitythreewavemixing}, instead of the four wave scheme, could allow a larger compression point. In addition, this would spectrally separate the signal and pump tones. Three wave mixing can readily be implemented in such gate tunable junctions by pumping through the gate. 

The use of semiconducting nanowires allows the integration of larger gap superconductors such as Sn \cite{sharmaSnInAsNanowireShadowDefined2025a} or Pb \cite{kanneEpitaxialPbInAs2021}, which would increase the critical current, improve the dynamic range, and allow operation of the device under magnetic field. 

A comparison of the nanowire JPA to other types of parametic amplifiers is provided in SM section VII \cite{Note1}. Our device is complementary to the kinetic inductance parametric amplifiers also based on hybrid nanowires but with a distributed nonlinearity \cite{splitthoffGatetunableKineticInductance2024,splitthoffGateTunableKineticInductance2022}.
With the recent developments in nanowire based superconducting qubits \cite{pita-vidalNovelQubitsHybrid2025}, it will be possible in the future to co-integrate, on the same chip, qubits and Josephson amplifiers made using the very same semiconductor-superconductor materials.

\begin{acknowledgements}
    
We thank Frederic Gay, Thomas Kanne and Laurent Cagnon for experimental assistance. The samples were prepared at the Nanofab clean room facility of the Néel Institute. We are grateful for the Chair of Excellence established by the Labex LANEF (CEA, CNRS and University of Grenoble). The work was also supported by the Carlsberg Foundation, and the Novo Nordisk Foundation SolidQ project, the French National Research Agency (ANR) in the framework of the Graphmon project (ANR-19-CE47-0007) and a French government grant managed by the ANR agency under the ‘France 2030 plan’ (ANR-22-PETQ-0003).  

\end{acknowledgements}

\section*{Supplemental Material}
Supplemental Material is available \cite{Note1}; 
detailed device fabrication procedure, device images, experimental measurement setup, measurement of current-phase relationship, noise analysis, comparisons to other amplifiers.


\section*{Appendices}
\appendix
\section{Microwave device design}\label{App:design}

The Josephson junction constituted by parallel InAs nanowires is contacted by two superconducting aluminum electrodes, embedded into a half-wavelength microwave resonator at the voltage node (Fig. \ref{fig:device_schematic}c). 
Electromagnetic simulations (\textit{Sonnet Software, Inc.}) yielded an estimate of the bare resonance frequency $\omega_0/2\pi$ = 6.5 GHz without the junction.
The microwave resonator is capacitively coupled to a 50 $\Omega$ transmission line. The geometry of the capacitor allows to tune the coupling quality factor $Q_c = \omega_r/\kappa_c$ and we targeted a value of 100. The internal quality factor $Q_i$ being typically around few thousands, ensures that the resonator is in the required over-coupled regime.
Two bias lines going directly to the Josephson junction (Fig. \ref{fig:device_schematic}c, ''DC lines'') enable dc transport measurements and a gate-line is added to tune the semiconductor (''Top-gate''). Large inductances were added to these lines to filter any unwanted microwave signal. 

\section{Device fabrication}

We grow InAs nanowires via the Vapor-Liquid-Solid (VLS) and Vapor-Solid (VS) mechanisms using Molecular Beam Epitaxy (MBE) following Ref. \cite{kanneDoubleNanowiresHybrid2022a}. We use lithographically-defined gold particles as catalysts.
Five or ten catalyst particles are arranged in a linear array allowing for the growth of parallel nanowires (Fig. \ref{fig:device_schematic}b). After loading the pre-patterned substrate inside the MBE chamber, it is annealed at 500\textdegree C for 5 min under As$_4$ overpressure, followed by an axial (VLS) growth at 450\textdegree C at In and As beam equivalent pressure (BEP) of 1.2E-7 Torr and 7.0E-6 Torr, respectively, for 80 min. An InAs shell is then grown at 350\textdegree C for 60 min around the nanowires (VS) in order to increase their diameter. This two step process has the advantage of growing long but sufficiently thick nanowires that form a dense array.
After cooling down the substrate to -50°C, the growth is followed by the deposition of a 35 nm Al half-shell without breaking the vacuum.

Parallel InAs nanowires are transferred onto an intrinsic silicon substrate and processed using e-beam lithography and positive resist. We deposit Ti/Al (5 nm/450 nm) using an e-beam evaporator to electrically contact the nanowires and form the microwave circuit (resonator, transmission line and dc lines). Ar$^+$ milling is used beforehand to remove the native aluminum oxide. The Josephson junction is formed by selectively etching part of the in situ grown aluminum using Transene aluminum etchant type D for 12 seconds at 50\textdegree C. We use atomic layer deposition to deposit around 10 nm of Hf$_2$O$_3$ as a top-gate dielectric. A final step consists of depositing the top-gate electrode using Ti/Al (5 nm/450 nm). A detailed description of the fabrication steps is available in SM section I \cite{Note1}.

\section{DC and microwave measurements}\label{App:meas}

We perform measurements in a wet dilution fridge at a base temperature of 25 mK. The dc transport measurements are done using standard lock-in techniques at a frequency of 17.17 Hz. The device JPA04 is measured in a quasi four-probe configuration while device JPA09 is measured in two-probes, so a constant line resistance of 1180 $\Omega$ is subtracted.
The microwave resonator is probed in reflection ($S_{11}$) using a Vector Network Analyzer (VNA). Power-dependent measurements are performed by setting a constant output power at the VNA and changing the attenuation using a variable attenuator. Another microwave source is used to send the pump tone through the same microwave port as the VNA using a beam splitter. The JPA output is amplified by a HEMT at 4~K. The setup is shown in the SM section II \cite{Note1}.

\bibliography{biblio_PR}

@article{samanta2026,
    author = {Samanta, Priyanka and Sarkar, Joydip and Samuel, Ashish Abhraham and Chand, Madhavi and Watanabe, Kenji and Taniguchi, Takashi and Deshmukh, Mandar M.},
    title = {Probing tunable Kerr nonlinearity in graphene Josephson junctions},
    journal = {Applied Physics Letters},
    volume = {129},
    number = {13},
    pages = {132603},
    year = {2026},
    month = {09},
    issn = {0003-6951},
    doi = {10.1063/5.0344221},
    url = {https://doi.org/10.1063/5.0344221},
}

@misc{buccheri2026kerrnonlinearitythreewavemixing,
      title={Kerr nonlinearity and three-wave mixing in superconducting resonators hosting Al-InAs weak links}, 
      author={Vittorio Buccheri and Ivo P. C. Cools and Nermin Trnjanin and Ankit Khola and Oleg Shvetsov and Thomas Kanne and Jesper Nygård and Attila Geresdi and Simone Gasparinetti},
      year={2026},
      eprint={2608.28428},
      archivePrefix={arXiv},
      primaryClass={quant-ph},
      url={https://arxiv.org/abs/2608.28428}, 
}

@article{blonderTransitionMetallicTunneling1982,
  title = {Transition from Metallic to Tunneling Regimes in Superconducting Microconstrictions: {{Excess}} Current, Charge Imbalance, and Supercurrent Conversion},
  shorttitle = {Transition from Metallic to Tunneling Regimes in Superconducting Microconstrictions},
  author = {Blonder, G. E. and Tinkham, M. and Klapwijk, T. M.},
  year = 1982,
  month = apr,
  journal = {Physical Review B},
  volume = {25},
  number = {7},
  pages = {4515--4532},
  issn = {0163-1829},
  doi = {10.1103/PhysRevB.25.4515},
  urldate = {2025-04-07},
  copyright = {http://link.aps.org/licenses/aps-default-license},
  langid = {english}
}

@article{boakninDispersiveMicrowaveBifurcation2007,
  title = {Dispersive Microwave Bifurcation of a Superconducting Resonator Cavity Incorporating a {{Josephson}} Junction},
  author = {Boaknin, E. and Manucharyan, V. E. and Fissette, S. and Metcalfe, M. and Frunzio, L. and Vijay, R. and Siddiqi, I. and Wallraff, A. and Schoelkopf, R. J. and Devoret, M.},
  year = 2007,
  month = feb,
  journal = {arXiv preprint},
  volume = {arXiv:cond-mat/0702445},
  eprint = {cond-mat/0702445},
  doi = {10.48550/arXiv.cond-mat/0702445},
  archiveprefix = {arXiv},
  langid = {english}
}

@article{bourassaJosephsonJunctionembeddedTransmissionline2012,
  title = {Josephson Junction-Embedded Transmission-Line Resonators: From {{Kerr}} Medium to in-Line Transmon},
  shorttitle = {Josephson Junction-Embedded Transmission-Line Resonators},
  author = {Bourassa, J. and Beaudoin, F. and Gambetta, Jay M. and Blais, A.},
  year = 2012,
  month = jul,
  journal = {Physical Review A},
  volume = {86},
  number = {1},
  eprint = {1204.2237},
  primaryclass = {cond-mat, physics:quant-ph},
  pages = {013814},
  issn = {1050-2947, 1094-1622},
  doi = {10.1103/PhysRevA.86.013814},
  urldate = {2023-03-01},
  archiveprefix = {arXiv},
  langid = {english}
}

@article{butseraenGatetunableGrapheneJosephson2022,
  title = {A Gate-Tunable Graphene {{Josephson}} Parametric Amplifier},
  author = {Butseraen, Guilliam and Ranadive, Arpit and Aparicio, Nicolas and Rafsanjani Amin, Kazi and Juyal, Abhishek and Esposito, Martina and Watanabe, Kenji and Taniguchi, Takashi and Roch, Nicolas and Lefloch, Fran{\c c}ois and Renard, Julien},
  year = 2022,
  month = nov,
  journal = {Nature Nanotechnology},
  volume = {17},
  number = {11},
  pages = {1153--1158},
  issn = {1748-3387, 1748-3395},
  doi = {10.1038/s41565-022-01235-9},
  urldate = {2023-01-10},
  langid = {english}
}

@article{casparisGatemonBenchmarkingTwoQubit2016,
  title = {Gatemon {{Benchmarking}} and {{Two-Qubit Operation}}},
  author = {Casparis, L. and Larsen, T. W. and Olsen, M. S. and Kuemmeth, F. and Krogstrup, P. and Nyg{\aa}rd, J. and Petersson, K. D. and Marcus, C. M.},
  year = 2016,
  month = apr,
  journal = {Physical Review Letters},
  volume = {116},
  number = {15},
  eprint = {1512.09195},
  primaryclass = {cond-mat, physics:quant-ph},
  pages = {150505},
  issn = {0031-9007, 1079-7114},
  doi = {10.1103/PhysRevLett.116.150505},
  urldate = {2024-01-15},
  archiveprefix = {arXiv},
  langid = {english}
}

@article{castellanos-beltranAmplificationSqueezingQuantum2008,
  title = {Amplification and Squeezing of Quantum Noise with a Tunable {{Josephson}} Metamaterial},
  author = {{Castellanos-Beltran}, M. A. and Irwin, K. D. and Hilton, G. C. and Vale, L. R. and Lehnert, K. W.},
  year = 2008,
  month = dec,
  journal = {Nature Physics},
  volume = {4},
  number = {12},
  pages = {929--931},
  issn = {1745-2473, 1745-2481},
  doi = {10.1038/nphys1090},
  urldate = {2024-08-09},
  copyright = {http://www.springer.com/tdm},
  langid = {english}
}

@article{castellanos-beltranBandwidthDynamicRange2009,
  title = {Bandwidth and {{Dynamic Range}} of a {{Widely Tunable Josephson Parametric Amplifier}}},
  author = {{Castellanos-Beltran}, M.A. and Irwin, K.D. and Vale, L.R. and Hilton, G.C. and Lehnert, K.W.},
  year = 2009,
  month = jun,
  journal = {IEEE Transactions on Applied Superconductivity},
  volume = {19},
  number = {3},
  pages = {944--947},
  issn = {1051-8223, 1558-2515},
  doi = {10.1109/TASC.2009.2018119},
  urldate = {2025-04-04},
  copyright = {https://ieeexplore.ieee.org/Xplorehelp/downloads/license-information/IEEE.html},
  langid = {english}
}

@article{castellanos-beltranWidelyTunableParametric2007,
  title = {A Widely Tunable Parametric Amplifier Based on a {{SQUID}} Array Resonator},
  author = {{Castellanos-Beltran}, M. A. and Lehnert, K. W.},
  year = 2007,
  month = aug,
  journal = {Applied Physics Letters},
  volume = {91},
  number = {8},
  eprint = {0706.2373},
  primaryclass = {cond-mat},
  pages = {083509},
  issn = {0003-6951, 1077-3118},
  doi = {10.1063/1.2773988},
  urldate = {2023-05-08},
  archiveprefix = {arXiv},
  langid = {english}
}

@article{cavesQuantumLimitsNoise1982,
  title = {Quantum Limits on Noise in Linear Amplifiers},
  author = {Caves, Carlton M.},
  year = 1982,
  month = oct,
  journal = {Physical Review D},
  volume = {26},
  number = {8},
  pages = {1817--1839},
  issn = {0556-2821},
  doi = {10.1103/PhysRevD.26.1817},
  urldate = {2023-03-22},
  langid = {english}
}

@article{cheungPhotonmediatedLongrangeCoupling2024,
  title = {Photon-Mediated Long-Range Coupling of Two {{Andreev}} Pair Qubits},
  author = {Cheung, L. Y. and Haller, R. and Kononov, A. and Ciaccia, C. and Ungerer, J. H. and Kanne, T. and Nyg{\aa}rd, J. and Winkel, P. and Reisinger, T. and Pop, I. M. and Baumgartner, A. and Sch{\"o}nenberger, C.},
  year = 2024,
  month = nov,
  journal = {Nature Physics},
  volume = {20},
  number = {11},
  pages = {1793--1797},
  issn = {1745-2473, 1745-2481},
  doi = {10.1038/s41567-024-02630-w},
  urldate = {2025-09-25},
  langid = {english}
}

@article{clerkIntroductionQuantumNoise2010,
  title = {Introduction to {{Quantum Noise}}, {{Measurement}} and {{Amplification}}},
  author = {Clerk, A. A. and Devoret, M. H. and Girvin, S. M. and Marquardt, F. and Schoelkopf, R. J.},
  year = 2010,
  month = apr,
  journal = {Reviews of Modern Physics},
  volume = {82},
  number = {2},
  eprint = {0810.4729},
  primaryclass = {cond-mat, physics:quant-ph},
  pages = {1155--1208},
  issn = {0034-6861, 1539-0756},
  doi = {10.1103/RevModPhys.82.1155},
  urldate = {2024-07-30},
  archiveprefix = {arXiv},
  langid = {english}
}

@article{eichlerControllingDynamicRange2014,
  title = {Controlling the Dynamic Range of a {{Josephson}} Parametric Amplifier},
  author = {Eichler, Christopher and Wallraff, Andreas},
  year = 2014,
  month = dec,
  journal = {EPJ Quantum Technology},
  volume = {1},
  number = {1},
  pages = {2},
  issn = {2196-0763},
  doi = {10.1140/epjqt2},
  urldate = {2024-02-24},
  langid = {english}
}

@article{fatemiNonlinearityTransparentSNS2025,
  title = {Nonlinearity of Transparent {{SNS}} Weak Links Decreases Sharply with Length},
  author = {Fatemi, Valla and Kurilovich, Pavel D. and Akhmerov, Anton R. and van Heck, Bernard},
  year = 2025,
  month = mar,
  journal = {SciPost Physics},
  volume = {18},
  number = {3},
  eprint = {2410.01913},
  primaryclass = {cond-mat},
  pages = {091},
  issn = {2542-4653},
  doi = {10.21468/SciPostPhys.18.3.091},
  urldate = {2025-09-29},
  archiveprefix = {arXiv},
  langid = {english}
}

@article{frattiniOptimizingNonlinearityDissipation2018,
  title = {Optimizing the {{Nonlinearity}} and {{Dissipation}} of a {{SNAIL Parametric Amplifier}} for {{Dynamic Range}}},
  author = {Frattini, N. E. and Sivak, V. V. and Lingenfelter, A. and Shankar, S. and Devoret, M. H.},
  year = 2018,
  month = nov,
  journal = {Physical Review Applied},
  volume = {10},
  number = {5},
  pages = {054020},
  issn = {2331-7019},
  doi = {10.1103/PhysRevApplied.10.054020},
  urldate = {2025-01-06},
  langid = {english}
}

@article{goffmanConductionChannelsInAsAl2017,
  title = {Conduction Channels of an {{InAs-Al}} Nanowire {{Josephson}} Weak Link},
  author = {Goffman, M F and Urbina, C and Pothier, H and Nyg{\aa}rd, J and Marcus, C M and Krogstrup, P},
  year = 2017,
  month = sep,
  journal = {New Journal of Physics},
  volume = {19},
  number = {9},
  pages = {092002},
  issn = {1367-2630},
  doi = {10.1088/1367-2630/aa7641},
  urldate = {2024-03-18},
  langid = {english}
}

@article{golubovCurrentphaseRelationJosephson2004,
  title = {The Current-Phase Relation in {{Josephson}} Junctions},
  author = {Golubov, A. A. and Kupriyanov, M. {\relax Yu}. and Il'ichev, E.},
  year = 2004,
  month = apr,
  journal = {Reviews of Modern Physics},
  volume = {76},
  number = {2},
  pages = {411--469},
  issn = {0034-6861, 1539-0756},
  doi = {10.1103/RevModPhys.76.411},
  urldate = {2024-11-18},
  copyright = {http://link.aps.org/licenses/aps-default-license},
  langid = {english}
}

@article{hallerPhasedependentMicrowaveResponse2022,
  title = {Phase-Dependent Microwave Response of a Graphene {{Josephson}} Junction},
  author = {Haller, Roy and F{\"u}l{\"o}p, Gerg{\H o} and Indolese, David and Ridderbos, Joost and Kraft, Rainer and Cheung, Luk Yi and Ungerer, Jann Hinnerk and Watanabe, Kenji and Taniguchi, Takashi and Beckmann, Detlef and Danneau, Romain and Virtanen, Pauli and Sch{\"o}nenberger, Christian},
  year = 2022,
  month = mar,
  journal = {Physical Review Research},
  volume = {4},
  number = {1},
  eprint = {2108.00989},
  primaryclass = {cond-mat},
  pages = {013198},
  issn = {2643-1564},
  doi = {10.1103/PhysRevResearch.4.013198},
  urldate = {2024-01-15},
  archiveprefix = {arXiv},
  langid = {english}
}

@article{haoKerrNonlinearityParametric2024,
  title = {Kerr Nonlinearity and Parametric Amplification with an {{Al-InAs}} Superconductor--Semiconductor {{Josephson}} Junction},
  author = {Hao, Z. and Shaw, T. and Hatefipour, M. and Strickland, W. M. and Elfeky, B. H. and Langone, D. and Shabani, J. and Shankar, S.},
  year = 2024,
  month = jun,
  journal = {Applied Physics Letters},
  volume = {124},
  number = {25},
  pages = {254003},
  issn = {0003-6951, 1077-3118},
  doi = {10.1063/5.0205053},
  urldate = {2025-05-15},
  langid = {english}
}

@article{haysCoherentManipulationAndreev2021,
  title = {Coherent Manipulation of an {{Andreev}} Spin Qubit},
  author = {Hays, M. and Fatemi, V. and Bouman, D. and Cerrillo, J. and Diamond, S. and Serniak, K. and Connolly, T. and Krogstrup, P. and Nyg{\aa}rd, J. and Levy Yeyati, A. and Geresdi, A. and Devoret, M. H.},
  year = 2021,
  month = jul,
  journal = {Science},
  volume = {373},
  number = {6553},
  pages = {430--433},
  issn = {0036-8075, 1095-9203},
  doi = {10.1126/science.abf0345},
  urldate = {2025-04-22},
  langid = {english}
}

@article{kamalSignaltopumpBackAction2009,
  title = {Signal-to-Pump Back Action and Self-Oscillation in Double-Pump {{Josephson}} Parametric Amplifier},
  author = {Kamal, Archana and Marblestone, Adam and Devoret, Michel},
  year = 2009,
  month = may,
  journal = {Physical Review B},
  volume = {79},
  number = {18},
  pages = {184301},
  issn = {1098-0121, 1550-235X},
  doi = {10.1103/PhysRevB.79.184301},
  urldate = {2024-07-29},
  copyright = {http://link.aps.org/licenses/aps-default-license},
  langid = {english}
}

@article{kanneDoubleNanowiresHybrid2022a,
  title = {Double {{Nanowires}} for {{Hybrid Quantum Devices}}},
  author = {Kanne, Thomas and Olsteins, Dags and Marnauza, Mikelis and Vekris, Alexandros and Estrada Salda{\~n}a, Juan Carlos and Loric, Sara and Schlosser, Rasmus D. and Ross, Daniel and Csonka, Szabolcs and Grove-Rasmussen, Kasper and Nyg{\aa}rd, Jesper},
  year = 2022,
  month = feb,
  journal = {Advanced Functional Materials},
  volume = {32},
  number = {9},
  pages = {2107926},
  issn = {1616-301X, 1616-3028},
  doi = {10.1002/adfm.202107926},
  urldate = {2023-01-12},
  langid = {english}
}

@article{kanneEpitaxialPbInAs2021,
  title = {Epitaxial {{Pb}} on {{InAs}} Nanowires},
  author = {Kanne, Thomas and Marnauza, Mikelis and Olsteins, Dags and Carrad, Damon J. and Sestoft, Joachim E. and {de Bruijckere}, Joeri and Zeng, Lunjie and Johnson, Erik and Olsson, Eva and {Grove-Rasmussen}, Kasper and Nyg{\aa}rd, Jesper},
  year = 2021,
  month = jul,
  journal = {Nature Nanotechnology},
  volume = {16},
  number = {7},
  eprint = {2002.11641},
  primaryclass = {cond-mat},
  pages = {776--781},
  issn = {1748-3387, 1748-3395},
  doi = {10.1038/s41565-021-00900-9},
  urldate = {2022-10-06},
  archiveprefix = {arXiv},
  langid = {english}
}

@article{kaufmanJosephsonParametricAmplifier2023,
  title = {Josephson Parametric Amplifier with {{Chebyshev}} Gain Profile and High Saturation},
  author = {Kaufman, Ryan and White, Theodore and Dykman, Mark I. and Iorio, Andrea and Stirling, George and Hong, Sabrina and Opremcak, Alex and Bengtsson, Andreas and Faoro, Lara and Bardin, Joseph C. and Burger, Tim and Gasca, Robert and Naaman, Ofer},
  year = 2023,
  month = may,
  journal = {arXiv preprint},
  volume = {arXiv.2305.17816},
  eprint = {2305.17816},
  primaryclass = {quant-ph},
  doi = {arXiv.2305.17816},
  archiveprefix = {arXiv},
  langid = {english}
}

@article{krogstrupEpitaxySemiconductorSuperconductor2015,
  title = {Epitaxy of Semiconductor--Superconductor Nanowires},
  author = {Krogstrup, P. and Ziino, N. L. B. and Chang, W. and Albrecht, S. M. and Madsen, M. H. and Johnson, E. and Nyg{\aa}rd, J. and Marcus, C.~M. and Jespersen, T. S.},
  year = 2015,
  month = apr,
  journal = {Nature Materials},
  volume = {14},
  number = {4},
  pages = {400--406},
  issn = {1476-1122, 1476-4660},
  doi = {10.1038/nmat4176},
  urldate = {2023-07-03},
  langid = {english}
}

@article{larsenSemiconductorNanowireBasedSuperconducting2015,
  title = {A {{Semiconductor Nanowire-Based Superconducting Qubit}}},
  author = {Larsen, T. W. and Petersson, K. D. and Kuemmeth, F. and Jespersen, T. S. and Krogstrup, P. and Nygard, J. and Marcus, C. M.},
  year = 2015,
  month = sep,
  journal = {Physical Review Letters},
  volume = {115},
  number = {12},
  eprint = {1503.08339},
  primaryclass = {cond-mat, physics:quant-ph},
  pages = {127001},
  issn = {0031-9007, 1079-7114},
  doi = {10.1103/PhysRevLett.115.127001},
  urldate = {2024-01-15},
  archiveprefix = {arXiv},
  langid = {english}
}

@article{macklinQuantumlimitedJosephsonTravelingwave2015,
  title = {A near--Quantum-Limited {{Josephson}} Traveling-Wave Parametric Amplifier},
  author = {Macklin, C. and O'Brien, K. and Hover, D. and Schwartz, M. E. and Bolkhovsky, V. and Zhang, X. and Oliver, W. D. and Siddiqi, I.},
  year = 2015,
  month = oct,
  journal = {Science},
  volume = {350},
  number = {6258},
  pages = {307--310},
  issn = {0036-8075, 1095-9203},
  doi = {10.1126/science.aaa8525},
  urldate = {2023-05-09},
  langid = {english}
}

@article{malnouOptimalOperationJosephson2018,
  title = {Optimal {{Operation}} of a {{Josephson Parametric Amplifier}} for {{Vacuum Squeezing}}},
  author = {Malnou, M. and Palken, D. A. and Vale, Leila R. and Hilton, Gene C. and Lehnert, K. W.},
  year = 2018,
  month = apr,
  journal = {Physical Review Applied},
  volume = {9},
  number = {4},
  pages = {044023},
  issn = {2331-7019},
  doi = {10.1103/PhysRevApplied.9.044023},
  urldate = {2026-03-05},
  langid = {english}
}

@article{manucharyanMicrowaveBifurcationJosephson2007,
  title = {Microwave Bifurcation of a {{Josephson}} Junction: {{Embedding-circuit}} Requirements},
  shorttitle = {Microwave Bifurcation of a {{Josephson}} Junction},
  author = {Manucharyan, V. E. and Boaknin, E. and Metcalfe, M. and Vijay, R. and Siddiqi, I. and Devoret, M.},
  year = 2007,
  month = jul,
  journal = {Physical Review B},
  volume = {76},
  number = {1},
  pages = {014524},
  issn = {1098-0121, 1550-235X},
  doi = {10.1103/PhysRevB.76.014524},
  urldate = {2024-07-05},
  copyright = {http://link.aps.org/licenses/aps-default-license},
  langid = {english}
}

@article{nishioSupercurrentInAsNanowires2011,
  title = {Supercurrent through {{InAs}} Nanowires with Highly Transparent Superconducting Contacts},
  author = {Nishio, Takahiro and Kozakai, Tatsuya and Amaha, Shinichi and Larsson, Marcus and Nilsson, Henrik A and Xu, H Q and Zhang, Guoqiang and Tateno, Kouta and Takayanagi, Hideaki and Ishibashi, Koji},
  year = 2011,
  month = nov,
  journal = {Nanotechnology},
  volume = {22},
  number = {44},
  pages = {445701},
  issn = {0957-4484, 1361-6528},
  doi = {10.1088/0957-4484/22/44/445701},
  urldate = {2025-04-08},
  langid = {english}
}

@article{phanGatetunableSuperconductorsemiconductorParametric2023,
  title = {Gate-Tunable, Superconductor-Semiconductor Parametric Amplifier},
  author = {Phan, D. and {Falthansl-Scheinecker}, P. and Mishra, U. and Strickland, W. M. and Langone, D. and Shabani, J. and Higginbotham, A. P.},
  year = 2023,
  month = jun,
  journal = {Physical Review Applied},
  volume = {19},
  number = {6},
  eprint = {2206.05746},
  primaryclass = {cond-mat, physics:quant-ph},
  pages = {064032},
  issn = {2331-7019},
  doi = {10.1103/PhysRevApplied.19.064032},
  urldate = {2024-07-29},
  archiveprefix = {arXiv},
  langid = {english}
}

@article{pita-vidalNovelQubitsHybrid2025,
  title = {Novel Qubits in Hybrid Semiconductor-Superconductor Nanostructures},
  author = {{Pita-Vidal}, Marta and Souto, Rub{\'e}n Seoane and Goswami, Srijit and Andersen, Christian Kraglund and Katsaros, Georgios and Shabani, Javad and Aguado, Ram{\'o}n},
  year = 2025,
  month = dec,
  volume = {arXiv.2512.23336},
  eprint = {2512.23336},
  primaryclass = {cond-mat},
  journal = {arXiv preprint},
  doi = {10.48550/arXiv.2512.23336},
  urldate = {2026-03-05},
  archiveprefix = {arXiv}
}

@article{pita-vidalStrongTunableCoupling2024,
  title = {Strong Tunable Coupling between Two Distant Superconducting Spin Qubits},
  author = {{Pita-Vidal}, Marta and Wesdorp, Jaap J. and Splitthoff, Lukas J. and Bargerbos, Arno and Liu, Yu and Kouwenhoven, Leo P. and Andersen, Christian Kraglund},
  year = 2024,
  month = jul,
  journal = {Nature Physics},
  volume = {20},
  number = {7},
  pages = {1158--1163},
  issn = {1745-2473, 1745-2481},
  doi = {10.1038/s41567-024-02497-x},
  urldate = {2025-09-25},
  langid = {english}
}

@article{planatPhotonicCrystalJosephsonTravelingWave2020,
  title = {Photonic-{{Crystal Josephson Traveling-Wave Parametric Amplifier}}},
  author = {Planat, Luca and Ranadive, Arpit and Dassonneville, R{\'e}my and Puertas Mart{\'i}nez, Javier and L{\'e}ger, S{\'e}bastien and Naud, C{\'e}cile and Buisson, Olivier and {Hasch-Guichard}, Wiebke and Basko, Denis M. and Roch, Nicolas},
  year = 2020,
  month = apr,
  journal = {Physical Review X},
  volume = {10},
  number = {2},
  pages = {021021},
  issn = {2160-3308},
  doi = {10.1103/PhysRevX.10.021021},
  urldate = {2024-10-29},
  langid = {english}
}

@article{pradaAndreevMajoranaBound2020,
  title = {From {{Andreev}} to {{Majorana}} Bound States in Hybrid Superconductor--Semiconductor Nanowires},
  author = {Prada, Elsa and {San-Jose}, Pablo and {de Moor}, Michiel W. A. and Geresdi, Attila and Lee, Eduardo J. H. and Klinovaja, Jelena and Loss, Daniel and Nyg{\aa}rd, Jesper and Aguado, Ram{\'o}n and Kouwenhoven, Leo P.},
  year = 2020,
  month = sep,
  journal = {Nature Reviews Physics},
  volume = {2},
  number = {10},
  pages = {575--594},
  issn = {2522-5820},
  doi = {10.1038/s42254-020-0228-y},
  urldate = {2022-12-20},
  langid = {english}
}

@article{probstEfficientRobustAnalysis2015,
  title = {Efficient and Robust Analysis of Complex Scattering Data under Noise in Microwave Resonators},
  author = {Probst, S. and Song, F. B. and Bushev, P. A. and Ustinov, A. V. and Weides, M.},
  year = 2015,
  month = feb,
  journal = {Review of Scientific Instruments},
  volume = {86},
  number = {2},
  pages = {024706},
  issn = {0034-6748, 1089-7623},
  doi = {10.1063/1.4907935},
  urldate = {2024-06-25},
  langid = {english}
}

@article{rhoadsImpactNonlinearityDegenerate2010,
  title = {The Impact of Nonlinearity on Degenerate Parametric Amplifiers},
  author = {Rhoads, Jeffrey F. and Shaw, Steven W.},
  year = 2010,
  month = jun,
  journal = {Applied Physics Letters},
  volume = {96},
  number = {23},
  pages = {234101},
  issn = {0003-6951, 1077-3118},
  doi = {10.1063/1.3446851},
  urldate = {2025-02-14},
  langid = {english}
}

@article{sarkarQuantumnoiselimitedMicrowaveAmplification2022,
  title = {Quantum-Noise-Limited Microwave Amplification Using a Graphene {{Josephson}} Junction},
  author = {Sarkar, Joydip and Salunkhe, Kishor V. and Mandal, Supriya and Ghatak, Subhamoy and Marchawala, Alisha H. and Das, Ipsita and Watanabe, Kenji and Taniguchi, Takashi and Vijay, R. and Deshmukh, Mandar M.},
  year = 2022,
  month = nov,
  journal = {Nature Nanotechnology},
  volume = {17},
  number = {11},
  pages = {1147--1152},
  issn = {1748-3387, 1748-3395},
  doi = {10.1038/s41565-022-01223-z},
  urldate = {2023-06-08},
  langid = {english}
}

@article{schmidtBallisticGrapheneSuperconducting2018,
  title = {A Ballistic Graphene Superconducting Microwave Circuit},
  author = {Schmidt, Felix E. and Jenkins, Mark D. and Watanabe, Kenji and Taniguchi, Takashi and Steele, Gary A.},
  year = 2018,
  month = oct,
  journal = {Nature Communications},
  volume = {9},
  number = {1},
  pages = {4069},
  issn = {2041-1723},
  doi = {10.1038/s41467-018-06595-2},
  urldate = {2025-01-29},
  langid = {english}
}

@article{schmidtProbingCurrentphaseRelation2020,
  title = {Probing the Current-Phase Relation of Graphene {{Josephson}} Junctions Using Microwave Measurements},
  author = {Schmidt, Felix E. and Jenkins, Mark D. and Watanabe, Kenji and Taniguchi, Takashi and Steele, Gary A.},
  year = 2020,
  month = jul,
  journal = {arXiv preprint},
  volume = {arXiv:2007.09795},
  eprint = {2007.09795},
  primaryclass = {cond-mat},
  doi = {arXiv:2007.09795},
  archiveprefix = {arXiv},
  langid = {english}
}

@article{sharmaSnInAsNanowireShadowDefined2025a,
  title = {Sn-{{InAs Nanowire Shadow-Defined Josephson Junctions}}},
  author = {Sharma, Amritesh and Chen, An-Hsi and Dempsey, Connor P. and Purkayastha, Amrita and Pendharkar, Mihir and Tan, Susheng and Palmstr{\o}m, Christopher J. and Frolov, Sergey M. and Hocevar, Mo{\"i}ra},
  year = 2025,
  month = aug,
  journal = {Nano Letters},
  volume = {25},
  number = {34},
  pages = {12869--12875},
  issn = {1530-6984, 1530-6992},
  doi = {10.1021/acs.nanolett.5c02410},
  urldate = {2025-09-24},
  copyright = {https://doi.org/10.15223/policy-029},
  langid = {english}
}

@article{spantonCurrentPhaseRelations2017,
  title = {Current--Phase Relations of Few-Mode {{InAs}} Nanowire {{Josephson}} Junctions},
  author = {Spanton, Eric M. and Deng, Mingtang and Vaitiek{\.e}nas, Saulius and Krogstrup, Peter and Nyg{\aa}rd, Jesper and Marcus, Charles M. and Moler, Kathryn A.},
  year = 2017,
  month = dec,
  journal = {Nature Physics},
  volume = {13},
  number = {12},
  pages = {1177--1181},
  issn = {1745-2473, 1745-2481},
  doi = {10.1038/nphys4224},
  urldate = {2024-11-18},
  langid = {english}
}

@article{splitthoffGateTunableKineticInductance2022,
  title = {Gate-{{Tunable Kinetic Inductance}} in {{Proximitized Nanowires}}},
  author = {Splitthoff, Lukas Johannes and Bargerbos, Arno and Gr{\"u}nhaupt, Lukas and {Pita-Vidal}, Marta and Wesdorp, Jaap Joachim and Liu, Yu and Kou, Angela and Andersen, Christian Kraglund and Van Heck, Bernard},
  year = 2022,
  month = aug,
  journal = {Physical Review Applied},
  volume = {18},
  number = {2},
  pages = {024074},
  issn = {2331-7019},
  doi = {10.1103/PhysRevApplied.18.024074},
  urldate = {2023-04-26},
  langid = {english}
}

@article{splitthoffGatetunableKineticInductance2024,
  title = {Gate-Tunable Kinetic Inductance Parametric Amplifier},
  author = {Splitthoff, Lukas Johannes and Wesdorp, Jaap Joachim and {Pita-Vidal}, Marta and Bargerbos, Arno and Liu, Yu and Andersen, Christian Kraglund},
  year = 2024,
  month = jan,
  journal = {Physical Review Applied},
  volume = {21},
  number = {1},
  pages = {014052},
  issn = {2331-7019},
  doi = {10.1103/PhysRevApplied.21.014052},
  urldate = {2026-02-25},
  langid = {english}
}

@article{teufelSidebandCoolingMicromechanical2011,
  title = {Sideband Cooling of Micromechanical Motion to the Quantum Ground State},
  author = {Teufel, J. D. and Donner, T. and Li, Dale and Harlow, J. W. and Allman, M. S. and Cicak, K. and Sirois, A. J. and Whittaker, J. D. and Lehnert, K. W. and Simmonds, R. W.},
  year = 2011,
  month = jul,
  journal = {Nature},
  volume = {475},
  number = {7356},
  pages = {359--363},
  issn = {0028-0836, 1476-4687},
  doi = {10.1038/nature10261},
  urldate = {2024-11-29},
  copyright = {http://www.springer.com/tdm},
  langid = {english}
}

@article{vijayInvitedReviewArticle2009,
  title = {Invited {{Review Article}}: {{The Josephson}} Bifurcation Amplifier},
  shorttitle = {Invited {{Review Article}}},
  author = {Vijay, R. and Devoret, M. H. and Siddiqi, I.},
  year = 2009,
  month = nov,
  journal = {Review of Scientific Instruments},
  volume = {80},
  number = {11},
  pages = {111101},
  issn = {0034-6748, 1089-7623},
  doi = {10.1063/1.3224703},
  urldate = {2024-09-05},
  langid = {english}
}

@article{walterRealizingRapidHighFidelity2017,
  title = {Realizing {{Rapid}}, {{High-Fidelity}}, {{Single-Shot Dispersive Readout}} of {{Superconducting Qubits}}},
  author = {Walter, T. and Kurpiers, P. and Gasparinetti, S. and Magnard, P. and Potocnik, A. and Salathe, Y. and Pechal, M. and Mondal, M. and Oppliger, M. and Eichler, C. and Wallraff, A.},
  year = 2017,
  month = may,
  journal = {Physical Review Applied},
  volume = {7},
  number = {5},
  eprint = {1701.06933},
  primaryclass = {quant-ph},
  pages = {054020},
  issn = {2331-7019},
  doi = {10.1103/PhysRevApplied.7.054020},
  urldate = {2024-11-25},
  archiveprefix = {arXiv},
  langid = {english}
}

@article{whiteReadoutQuantumProcessor2023,
  title = {Readout of a Quantum Processor with High Dynamic Range {{Josephson}} Parametric Amplifiers},
  author = {White, T. C. and Opremcak, Alex and Sterling, George and Korotkov, Alexander and Sank, Daniel and Acharya, Rajeev and Ansmann, Markus and Arute, Frank and Arya, Kunal and Bardin, Joseph C. and Bengtsson, Andreas and Bourassa, Alexandre and Bovaird, Jenna and Brill, Leon and Buckley, Bob B. and Buell, David A. and Burger, Tim and Burkett, Brian and Bushnell, Nicholas and Chen, Zijun and Chiaro, Ben and Cogan, Josh and Collins, Roberto and Crook, Alexander L. and Curtin, Ben and Demura, Sean and Dunsworth, Andrew and Erickson, Catherine and Fatemi, Reza and {Flores-Burgos}, Leslie and Forati, Ebrahim and Foxen, Brooks and Giang, William and Giustina, Marissa and Dau, Alejandro Grajales and Hamilton, Michael C. and Harrington, Sean D. and Hilton, Jeremy and Hoffmann, Markus and Hong, Sabrina and Huang, Trent and Huff, Ashley and Iveland, Justin and Jeffrey, Evan and Kieferov{\'a}, M{\'a}rika and Kim, Seon and Klimov, Paul V. and Kostritsa, Fedor and Kreikebaum, John Mark and Landhuis, David and Laptev, Pavel and Laws, Lily and Lee, Kenny and Lester, Brian J. and Lill, Alexander and Liu, Wayne and Locharla, Aditya and Lucero, Erik and McCourt, Trevor and McEwen, Matt and Mi, Xiao and Miao, Kevin C. and Montazeri, Shirin and Morvan, Alexis and Neeley, Matthew and Neill, Charles and Nersisyan, Ani and Ng, Jiun How and Nguyen, Anthony and Nguyen, Murray and Potter, Rebecca and Quintana, Chris and Roushan, Pedram and Sankaragomathi, Kannan and Satzinger, Kevin J. and Schuster, Christopher and Shearn, Michael J. and Shorter, Aaron and Shvarts, Vladimir and Skruzny, Jindra and Smith, W. Clarke and Szalay, Marco and Torres, Alfredo and Woo, Bryan and Yao, Z. Jamie and Yeh, Ping and Yoo, Juhwan and Young, Grayson and Zhu, Ningfeng and Zobrist, Nicholas and Chen, Yu and Megrant, Anthony and Kelly, Julian and Naaman, Ofer},
  year = 2023,
  month = jan,
  journal = {Applied Physics Letters},
  volume = {122},
  number = {1},
  eprint = {2209.07757},
  primaryclass = {quant-ph},
  pages = {014001},
  issn = {0003-6951, 1077-3118},
  doi = {10.1063/5.0127375},
  urldate = {2025-05-13},
  archiveprefix = {arXiv},
  langid = {english}
}

@article{yamamotoFluxdrivenJosephsonParametric2008a,
  title = {Flux-Driven {{Josephson}} Parametric Amplifier},
  author = {Yamamoto, T. and Inomata, K. and Watanabe, M. and Matsuba, K. and Miyazaki, T. and Oliver, W. D. and Nakamura, Y. and Tsai, J. S.},
  year = 2008,
  month = jul,
  journal = {Applied Physics Letters},
  volume = {93},
  number = {4},
  pages = {042510},
  issn = {0003-6951, 1077-3118},
  doi = {10.1063/1.2964182},
  urldate = {2024-09-06},
  langid = {english}
}

@article{yangProximitizedJosephsonJunctions2021,
  title = {Proximitized {{Josephson}} Junctions in Highly-Doped {{InAs}} Nanowires Robust to Optical Illumination},
  author = {Yang, Lily and Steinhauer, Stephan and Strambini, Elia and Lettner, Thomas and Schweickert, Lucas and Versteegh, Marijn A M and Zannier, Valentina and Sorba, Lucia and Solenov, Dmitry and Giazotto, Francesco},
  year = 2021,
  month = feb,
  journal = {Nanotechnology},
  volume = {32},
  number = {7},
  pages = {075001},
  issn = {0957-4484, 1361-6528},
  doi = {10.1088/1361-6528/abc44e},
  urldate = {2025-04-08},
  langid = {english}
}

@article{yurkePerformanceCavityparametricAmplifiers2006,
  title = {Performance of Cavity-Parametric Amplifiers, Employing {{Kerr}} Nonlinearites, in the Presence of Two-Photon Loss},
  author = {Yurke, Bernard and Buks, Eyal},
  year = 2006,
  month = dec,
  journal = {Journal of Lightwave Technology},
  volume = {24},
  number = {12},
  eprint = {quant-ph/0505018},
  pages = {5054--5066},
  issn = {0733-8724},
  doi = {10.1109/JLT.2006.884490},
  urldate = {2024-07-29},
  archiveprefix = {arXiv},
  langid = {english}
}

@article{zellekensHardGapSpectroscopySelfDefined2020,
  title = {Hard-{{Gap Spectroscopy}} in a {{Self-Defined Mesoscopic In As}} / {{Al Nanowire Josephson Junction}}},
  author = {Zellekens, Patrick and Deacon, Russell and Perla, Pujitha and Fonseka, H. Aruni and M{\"o}rstedt, Timm and Hindmarsh, Steven A. and Bennemann, Benjamin and Lentz, Florian and Lepsa, Mihail I. and Sanchez, Ana M. and Gr{\"u}tzmacher, Detlev and Ishibashi, Koji and Sch{\"a}pers, Thomas},
  year = 2020,
  month = nov,
  journal = {Physical Review Applied},
  volume = {14},
  number = {5},
  pages = {054019},
  issn = {2331-7019},
  doi = {10.1103/PhysRevApplied.14.054019},
  urldate = {2025-07-31},
  langid = {english}
}

@article{zimmerParametricAmplificationMicrowaves1967,
  title = {Parametric Amplification of Microwaves in Superconducting Josephson Tunnel Junctions},
  author = {Zimmer, H.},
  year = 1967,
  month = apr,
  journal = {Applied Physics Letters},
  volume = {10},
  number = {7},
  pages = {193--195},
  issn = {0003-6951, 1077-3118},
  doi = {10.1063/1.1754906},
  urldate = {2023-02-15},
  langid = {english}
}

\end{document}